\documentclass[twocolumn,showpacs,preprintnumbers,amsmath,amssymb]{revtex4}
\usepackage{epsfig}
\usepackage{subfigure}
\usepackage{amsmath}
\usepackage{amssymb}
\usepackage{graphicx}
\usepackage{dcolumn}
\usepackage{bm}
\input epsf

\begin{document}

\title{Spatial patterns of desynchronization bursts in networks}

\author{Juan G. Restrepo}
\email{juanga@math.umd.edu}
\affiliation{
Institute for Research in Electronics and Applied Physics and Department of Mathematics, 
University of Maryland, College
Park, Maryland 20742
}
\author{Edward Ott}
\affiliation{
Institute for Research in Electronics and Applied Physics, Department of Physics
and Deparment of Electrical and Computer Engineering, University of Maryland, College
Park, Maryland 20742
}
\author{Brian R. Hunt}
\affiliation{
Institute for Physical Science and Technology and Department of Mathematics,
University of Maryland, College Park, Maryland 20742
}

\date{\today}

\begin{abstract}
We adapt a previous  model and analysis method (the {\it master stability function}),
extensively used for 
studying the stability of the synchronous state of networks of identical chaotic oscillators,
to the case of oscillators that are similar but not exactly identical.
We find that bubbling induced desynchronization bursts occur for some parameter values. These
bursts have spatial patterns, which can be predicted from the 
network connectivity matrix and the unstable periodic orbits embedded in the attractor.
We test the analysis of bursts by comparison with numerical experiments. 
In the case that no bursting occurs, we discuss the deviations from the exactly synchronous state caused
by the mismatch between oscillators.
\end{abstract}


\maketitle

\section{Introduction}

In this paper we study the synchronization of networks of
coupled chaotic units that are nearly, but not exactly, identical. In particular, we will
be concerned with the spatial patterns of desynchronization bursts
that appear when this synchronization is present but intermittent.

When two or more identical dynamical systems are coupled, they can
synchronize under appropriate circumstances. The synchronization
of chaotic units has been studied extensively \cite{pecora1,pikovsky} and is of
significance in biology \cite{elson}-\cite{mosekilde}, laser physics
\cite{roy}-\cite{uchida}, and
other areas \cite{hudson,cuomo}. At the same time, the importance of complex
networks has been recently appreciated, and progress has been made
towards their understanding, including characteristics that
might help distinguish qualitatively different networks \cite{newman1}-\cite{newman2}.
The dynamics of a network of coupled oscillators, and, in particular,
its synchronization, has therefore emerged as a subject of great
interest.

Pecora and Carroll \cite{pecora2} have proposed a model and analysis method
(the {\it master stability function}) for the study of the
stability of the synchronous state of networks of {\it identical}
coupled chaotic units, and this technique has recently been
extensively applied \cite{pecora3,takashi} to study the synchronization properties of
different kinds of networks of identical noiseless chaotic
units. These networks include small world \cite{smallworld} and scale-free
networks \cite{barabasi}.

The analysis of network synchronization by use of the master
stability function technique has so far assumed all the units to
be identical and noise-free, so that an exact synchronized state is possible. In
practice, however, even if one strives to make the oscillators the
same, they are still expected to have a small amount of parameter mismatch,
and a small amount of noise is also expected to be present. Under
such circumstances, it is known that the synchronization can be
interrupted by sporadic periods of desynchronization (bursts). The
bursts are typically caused by a periodic orbit that is embedded
in the synchronized chaotic  attractor and is unstable in a
direction transverse to the synchronization manifold. This
phenomenon is commonly referred to as {\it bubbling} \cite{ashwin}-\cite{bhott2}, 
and has been
studied extensively for two coupled oscillators \cite{rulkov,pecora4}.

Our purpose in this paper is to study desynchronization bursts in
networks of coupled chaotic nonidentical units. 
(Noise has a similar effect but will 
not be treated in this paper.) We will use the
master stability function approach and, in order to account for
the possibility of bubbling, we will also extend this approach to include
the stability of
embedded periodic orbits. In this case, the bursts have the added
feature of having spatial patterns on the network, and we find
that these patterns can be predicted from the network connectivity
matrix. 
We will show how these
bursts affect different parts of the network in different ways. In
particular, we will see how adding connections in a ring can
destabilize precisely those nodes that are the most connected,
leaving other parts of the network substantially synchronized.
(This a somewhat counterintuitive effect related to the fact that,
in some cases, increasing the coupling strength destabilizes the
synchronous state \cite{pecora2},\cite{pecora5}.)

Arbitrarily small amounts of mismatch will eventually, through the
bubbling mechanism, induce desynchronization bursts. We will show that
some of the spatial patterns of this possibly microscopic mismatch
might get amplified to a macroscopic size in the bursts. 
We will discuss how one can use knowledge of the parameter mismatch of
the dynamical units in the network to decrease the effective size of 
the mismatch driving the bursts, thereby improving the robustness
of the synchronization.

If synchronization is desired, the network and the parameters should be constructed
so that the synchronous state for the identical oscillator system 
is robustly stable (this implies the absence of 
noise or mismatch induced desynchronization bursts). Even then, the synchronization will
not be perfect if the oscillators have parameter mismatch. 
We will describe the characteristics of the deviations from exact synchronization 
in terms of the mismatch and the master stability function. 

This paper is organized as follows. In Sec.~II we review the
master stability function approach and apply it to the case of
coupled R\"{o}ssler units. We also discuss the bubbling mechanism
by including the embedded periodic orbits in the master stability
function analysis. In Sec.~III we numerically consider particular
networks as examples and show the resulting bursts and their
spatial patterns. The patterns we obtain are long and short
wavelength modes in a ring and localized bursts produced by 
strengthening of a single connection in a ring. 
In Sec.~IV we study the effects of the spatial patterns of the mismatch 
in the development of the bursts. In Sec.~V we study the 
deviations from the synchronous state caused by the mismatch 
when the synchronous state of the identical oscillator system is stable. 
In Sec.~VI we summarize our conclusions.

\section{Master stability function and bubbling}

We now briefly review the master stability function approach
introduced in \cite{pecora2}. Consider a system of $N$ dynamical units,  each
one of which, when isolated, satisfies 
 $\dot{X}_{i}  = F(X_{i},\mu_{i})$, where $i =
1,2,\dots N$, and $X_{i}$ is the $d$-dimensional state vector for unit $i$. 
In \cite{pecora2} the parameter vectors $\mu_{i}$ are 
taken to be the
same, $\mu_{i}=\mu$. Here, however, the parameter vectors $\mu_{i}$ are
in general different for each unit, but we assume the difference,
or {\it mismatch}, between them to be small. Generalizing the
situation treated in Ref.~\cite{pecora2} to the case where the individual
units are not identical (i.e., the $\mu_{i}$ are not all equal),
the system of coupled dynamical units is taken to be of the form

\begin{equation}\label{eq:coupled}
\dot{X}_{i} = F(X_{i},\mu_{i}) - g \sum_{j = 1}^{N}
G_{ij}H(X_{j}),
\end{equation}
where the coupling function $H$ is
independent of $i$ and $j$, and the matrix $G$ is a Laplacian matrix
($\sum_{j} G_{ij} = 0$) describing the topology of network
connections. For $i \neq j$, the entry $G_{ij}$ is zero if
oscillator $i$ is not connected to oscillator $j$ and nonzero
otherwise. The nondiagonal entries of $G$ are determined by the
connections, and the diagonal elements are the negative of the sum
of the nondiagonal matrix elements in their row. The coupling
constant $g$ determines the global strength of the coupling.

Assume first that all the dynamical units are identical, that is,
$\mu_{i} = \mu$. We will refer to this situation as the {\it
idealized} case. In this case there is an exactly synchronized
solution $X_{1} = X_{2} = \dots = X_{N} = s(t)$ whose time evolution is the same
as the uncoupled dynamics of a single unit, $\dot{s} = F(s,\mu)$. This
convenient result arises because the Pecora-Carroll model uses the
particular choice of coupling in (\ref{eq:coupled}) that ensures 
that the summation is identically zero when all of the $X_j$ are equal. 
We will denote this synchronization
manifold, $X_{1} = X_{2} = \dots = X_{N}$, by $M$. This manifold
is a $d$~-~dimensional surface within the $Nd$ - dimensional phase
space of Eq. (\ref{eq:coupled}).

The stability of the synchronized state can be determined from
the variational equations obtained by considering an infinitesimal
perturbation $\epsilon_{i}$ from the synchronous state, $X_{i}(t)
= s(t) + \epsilon_{i}(t)$,

\begin{equation}\label{eq:linearized}
\dot{\epsilon}_{i} = DF(s)\epsilon_{i} - g \sum_{j = 1}^{N}
G_{ij}DH(s)\epsilon_{j}.
\end{equation}
Let $\epsilon = [\epsilon_{1},\epsilon_{2},\dots,\epsilon_{N}]$
be the $d \times N$ matrix representing the deviation of the
entire network from the synchronized state. In matrix notation,
Eq.~(\ref{eq:linearized}) becomes

\begin{equation}\label{eq:matrixform}
\dot{\epsilon} = DF(s)\epsilon - g DH(s)\epsilon G^{T}.
\end{equation}
While (\ref{eq:matrixform}) allows for nonsymmetric coupling, 
we henceforth assume the coupling matrix $G$ to be symmetric,
$G = G^{T}$.
We  write
the symmetric matrix $G$ as $G=L\Lambda L^{T}$, where $\Lambda$ is
the diagonal matrix of real eigenvalues $\lambda_{1}, \lambda_{2},
\dots, \lambda_{N}$ of $G$ and $L$ is the orthogonal matrix whose
columns are the corresponding real orthonormal eigenvectors of $G$
($L^{T}L = I$). Define the $d\times N$ matrix $\eta =
[\eta_{1},\eta_{2},\dots,\eta_{N}]$ by $\epsilon = \eta L^{T}$.
Then Eq.~(\ref{eq:matrixform}) is equivalent to

\begin{equation}
\dot{\eta} = DF(s)\eta - g DH(s)\eta\Lambda.
\end{equation}
Componentwise,

\begin{equation}\label{eq:componentwise}
\dot{\eta}_{k} = \left(DF(s) - g \lambda_{k} DH(s)\right)\eta_{k}.
\end{equation}
The quantity $\eta_{k}$ is the weight of the $k^{th}$ eigenvector
of $G$ in the perturbation $\epsilon$.  The linear stability of
each `spatial' mode $k$ is determined by the stability of the zero
solution of (\ref{eq:componentwise}). As a consequence of the
condition $\sum_{j} G_{ij} = 0$, there is a special eigenvalue,
$\lambda = 0$, whose  eigenvector is $\epsilon_{N}
= [1,1,1,\dots,1]$, corresponding to perturbations {\it in} the
synchronization manifold $M$. Since these are not perturbations
from the synchronous state, the analysis is focused on the
perturbations corresponding to nonzero eigenvalues.

By introducing a scalar variable $\alpha = g\lambda_{k}$, the set
of equations given by (\ref{eq:componentwise}) can be encapsulated
in the single equation,

\begin{equation}\label{eq:master}
\dot{\eta} = \left(DF(s) - \alpha DH(s)\right)\eta.
\end{equation}
The {\it master stability  function} $\Psi (\alpha)$ \cite{pecora2} is the
largest Lyapunov exponent for this equation for a typical
trajectory in the attractor. This function depends only on the
coupling function $H$ and the chaotic dynamics of an individual
uncoupled element, but not on the network connectivity. The
network connectivity determines the eigenvalues $\lambda_{k}$
(independent of details of the dynamics of the chaotic units). In
the sense of typical Lyapunov exponents, the stability of the
synchronized  state of the network is determined by $\Psi_{*} =
\sup_{k} \Psi(g \lambda_{k})$, where $\Psi_{*}>0$ indicates
instability. Thus the Pecora-Carroll model cleanly breaks the
stability problem into two components, one from the dynamics
(obtaining $\Psi(\alpha)$) and one from the network (determining
the eigenvalues $\lambda_{k}$).

In contrast to previous work using the master stability function
technique, in this paper we are interested in the dynamics of
systems in which a small parameter mismatch is present. (Even
though in this paper our examples are restricted to the case of
mismatch, we emphasize that the same type of bursting phenomenon
is expected for identical oscillators if noise is present 
\cite{ashwin}-\cite{rulkov}.)
Although the synchronization manifold $M$  present in the dynamics
of the idealized system is, in general, not invariant for the 
system with mismatch,
it still may provide a useful approximation to the dynamics in
systems with small mismatch. If $M$ is stable for the idealized system, 
and the mismatch is small enough, then trajectories near $M$ will tend to stay 
near $M$, and we regard the vicinity of $M$ to be the
``synchronized'' state.
However, stability of $M$ in the idealized case of identical oscillators 
is not sufficient to guarantee robust synchronization in a real system where the oscillators 
are not identical\cite{ashwin}-\cite{rulkov}. 
While in  the vicinity of the
synchronization manifold $M$, a typical trajectory 
will eventually follow very closely a
periodic orbit embedded in the attractor of the idealized system. 
Some of these periodic
orbits may be unstable in a direction transverse to $M$. When in the
vicinity of a transversally unstable periodic orbit, mismatch (or
noise) will cause the trajectory have a component in the direction transverse to $M$
and hence to leave the vicinity of the
synchronization manifold $M$. If there are
no other attractors, the trajectory will eventually return to the
vicinity of $M$, and the process will repeat, the result being
bursts of desynchronization sporadically interrupting long
intervals of near synchronization. This type of dymanics is called 
bubbling \cite{ashwin}.

Thus, in the presence of mismatch (or noise), to determine the robustness of 
synchronization, it is necessary to
determine the transverse stability of the embedded periodic
orbits for the noiseless system of identical oscillators. For coupling as in (\ref{eq:coupled}), this
analysis is independent of the network, and such analyses have
been carried out before, e.g., for the analysis of {\it two}
coupled oscillators in Ref.~\cite{pecora4}. Equation (\ref{eq:master}) can be
used as before to construct the master stability function for each
periodic orbit, if the appropriate periodic trajectories are
inserted for $s(t)$ in (\ref{eq:matrixform}).

\begin{figure}[h]
\begin{center}
\epsfig{file = 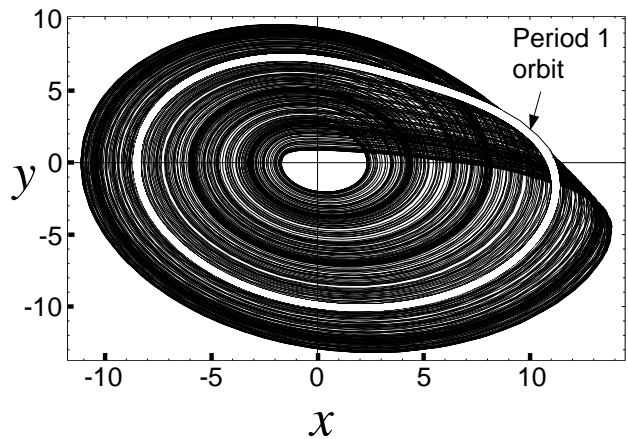, clip =  ,width=1\linewidth }
\caption{R\"{o}ssler attractor (projection onto $x-y$ plane) and
embedded period $1$ orbit, displayed as a thick white curve inside the 
attractor. 
The parameters are $a = b = 0.2$, $c=7$.} 
\label{fig:ros}
\end{center}
\end{figure}

\begin{figure}[h]
\begin{center}
\epsfig{file = 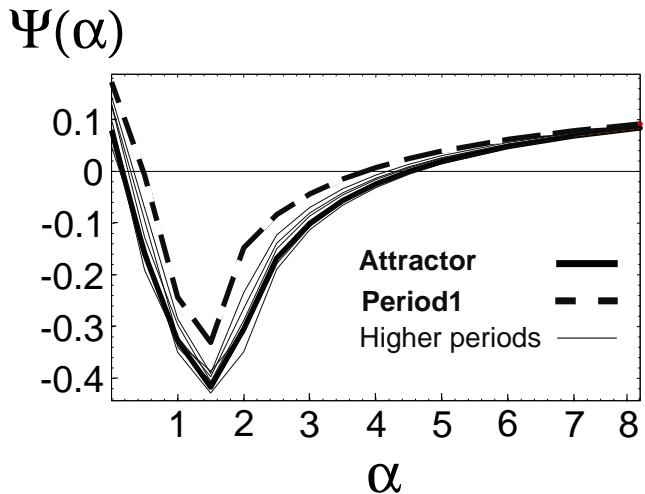, clip =  ,width=1\linewidth }
\caption{Master stability function $\Psi(\alpha)$ for a  typical
trajectory in the attractor (thick continuous curve), for the period
$1$ orbit (thick dashed curve), and for periodic orbits up to period
$4$ (thin curves). The curves for the four period $5$
orbits are similar to the latter and were left out for clarity.}
\label{fig:msf}
\end{center}
\end{figure}

As an example, in this paper we work with the R\"{o}ssler system \cite{rossler}:
\begin{eqnarray}\label{eq:rossler}
\dot{x} = - (y+z),\\\nonumber
\dot{y} = x + a y,\\\nonumber
\dot{z} = b + z(x-c).\\\nonumber
\end{eqnarray}
In terms of our previous notation, $d = 3$, $\mu=[a,b,c]^{T}$, and $X = [x,y,z]^{T}$.
We choose the parameters of the idealized system to be $a = b = 0.2$, $c =
7$. For these parameters, the system has a chaotic attractor (see
Fig.~\ref{fig:ros}).
We found the periodic orbits embedded in this attractor up to period five,
and performed the analysis described above on them.
We found these orbits by looking at the Poincare surface of 
section  $\{y =
0, x < 0\}$. To a good approximation, in this
surface of section the dynamics is well described by a one
dimensional map $x_{n+1}=f(x_{n})$, which we approximated using a
polynomial fit. From this approximation to $f$, we determined
periodic orbits of period $p$ by using Newton's method to find the
roots of $x = f^{p}(x)$, where $f^{p}$ denotes the $p$ times
composition of $f$. We found one period $1$ orbit, one period $2$
orbit, two period $3$ orbits, three period $4$ orbits, and four
period $5$ orbits. Using coupling through the $x$ coordinate,
\begin{equation}\label{eq:H}
H([x,y,z]^{T}) = [x,0,0]^{T},
\end{equation}
we obtained a stability function $\Psi(\alpha)$
for each orbit, the largest of which will determine if the
synchronization is robust.  Results are shown in
Fig.~\ref{fig:msf}. For all values of $\alpha$, we found that the
master stability function corresponding to the period $1$ orbit
(thick dashed curve) is larger than that for a typical chaotic orbit
(thick continuous curve), as well as those for the other periodic orbits we have
found (several of which are shown as thin curves).

Based on the discussion above, bubbling induced bursting should occur
whenever the master stability function for a typical  chaotic
orbit in the attractor is negative for $\alpha = g \lambda_{k}$
and all $k$, while the period one orbit has positive master
stability function for $\alpha = g \lambda_{k}$ for some value of
$k$. Denoting the master stability function for a typical chaotic orbit
by $\Psi_{0}(\alpha)$ (Thick continuous curve in Fig.~\ref{fig:msf}) and for the 
period one orbit by $\Psi_{1}(\alpha)$ (Thick dashed curve in Fig.~\ref{fig:msf}),
the {\it bubbling region} of $\alpha$ corresponds to $\Psi_{0}(\alpha)<0$,
$\Psi_{1}(\alpha)>0$. In our example, this region corresponds to 
$0.16 < \alpha < 0.48$ or $3.8 < \alpha < 4.5$. The range $0.48 < \alpha
< 3.8$ will be referred to as the {\it stable region}, and the remaining zone will be
called the {\it unstable region}.

If a network of slightly mismatched chaotic systems coupled according 
to \ref{eq:coupled}is to be
robustly synchronizable  without bursts of desynchronization, $g\lambda_k$ must lie 
in the stable region for all $k$, where $\lambda_k$ is the $k$th eigenvalue of $G$. If
$g\lambda_k$ lies in the stable region for some $k$ and in the bubbling region for other
$k$, then bubbling will typically occur.

\section{Examples}

In this Section we provide examples of spatially patterned
bursting by considering different configurations of the chaotic
units. We will first work with the units connected in a ring with
each connection of equal strength. The Laplacian matrix $G$ for
this arrangement is

\begin{equation}\label{eq:laplaciandots}
G = \left( \begin{array}{ccccccc}
     2 & -1 & 0 & 0 & \cdots & 0 & -1 \\
     -1 & 2 & -1 & 0 & \cdots & 0 & 0  \\
     0 &  -1 & 2 & -1  & \cdots & 0 & 0  \\
     \vdots&\vdots&\vdots&\vdots&\vdots&\vdots&\vdots\\
     -1 & 0 & \cdots & 0 & 0 & -1 & 2
\end{array} \right),
\end{equation}
and its eigenvalues are given by $\lambda_{k}= 4\sin^{2}(\frac{\pi
k}{N})$.  Since $\lambda_{k} = \lambda_{N-k}$,
each eigenvalue has multiplicity two, with the exception of $\lambda_{N} = 0$, and,
if $N$ is even, $\lambda_{\frac{N}{2}} = 4$.
The matrix $G$ is {\it shift invariant}, that is, its entries
satisfy, modulo $N$, $G_{i, j} = G_{0 , i-j}$.
Under these conditions, the diagonalization procedure described above
corresponds to a discrete Fourier transform \cite{pecora5}. For the eigenvalue
$\lambda_{k}$ we choose the eigenvector $w_{k}$ given by  
$w_{k}\propto [\sin(\frac{2\pi j k}{N})]_{j=1}^{N}$ for $1\leq k < \frac{N}{2}$, and 
by $w_{k}\propto [\cos(\frac{2\pi j k}{N})]_{j=1}^{N}$  for $\frac{N}{2}\leq k \leq N$. 
(Due to the degeneracy of the eigenvalues in this case, there is some
arbitrariness in choosing the eigenvectors.)
Thus, the longest wavelength modes have the smallest eigenvalues, and
viceversa.

\subsection{Long wavelength burst}

First we consider a case in which bursting of the longest
wavelength  mode occurs. We consider $N = 12$ and $g = 0.71$. With
these values, the longest wavelength mode corresponds to $\alpha =
g \lambda_{1} \approx 0.19$. This value is in the bubbling region,
and all other modes are in the stable region.

To introduce heterogeneity in the dynamical units, we imagine that
we have mismatch predominantly in one of the parameters, say $a$. 
We simulate this mismatch
by adding random perturbations to the parameter $a$ of each oscillator. 
These perturbations are uniformly
distributed within  a $\pm 0.5\%$ range; i.e., $a_{i}$ is chosen randomly 
in the interval $[0.995a,1.005a]$, where
$a$ is the parameter value of the unperturbed system
($a = 0.2$). The parameters $b$ and $c$ were taken to be the same for each 
oscillator, $b_{i} = b = 0.2$, $c_{i} = c = 7$. How a particular 
choice of the mismatch affects the 
bubbling process will be discussed in Section~IV.

\begin{figure}[h]
\begin{center}
\epsfig{file = 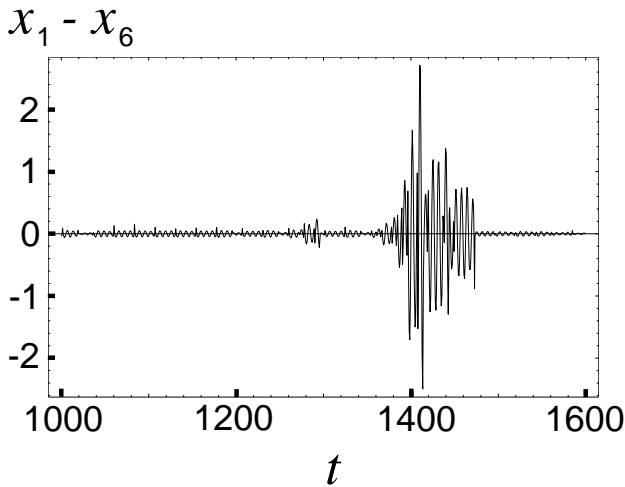, clip =  ,width=1.0\linewidth }
      
\caption{$x_{1}-x_{6}$ as a function of time for $N = 12$ R\"ossler systems
                connected in a ring with $g = 0.71$. Note the desynchronization
                burst which starts at $t\approx 1380$.}
\label{fig:burst}
\end{center}
\end{figure}

We solved the $12$ coupled differential
equations (Eq.~(\ref{eq:coupled})) with the initial conditions chosen 
near the attractor in
the synchronization manifold. In Fig.~\ref{fig:burst} we plot the
quantity $x_{1}-x_{6}$ for $1000\leq t \leq 1600$.
Most of the time, this variable is close to zero, as  expected if
the oscillators are synchronized. Approximately at the time $t =
1380$, this difference grows, reaching magnitudes close to $3$.
By time $t = 1500$, the difference has decreased and is again
close to zero. 
\begin{figure}[h]
\begin{center}
\epsfig{file = 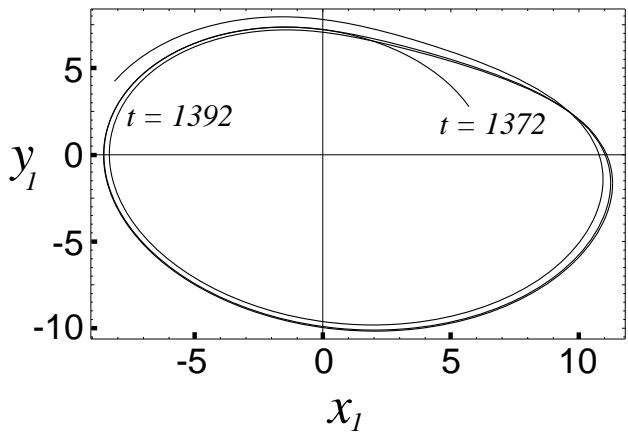, clip =  ,width=1.0\linewidth }
\caption{ $x_{1}$ versus $y_{1}$ for $1372 \leq t \leq 1392$.
During  this period, which corresponds approximately to the
starting point of the burst in Fig.~\ref{fig:burst}, the
trajectory follows closely the transversally unstable period $1$
orbit embedded in the attractor (See Fig.~\ref{fig:ros}).
} \label{fig:p1}
\end{center}
\end{figure}

To confirm the mediating role of the
embedded unstable periodic orbits in the development of the
desynchronization burst, we show in Fig.~\ref{fig:p1} a plot of
$x_{1}$ versus $y_{1}$ from $t = 1372$ to $t = 1392$, which is
near the start of the burst. During this time, the trajectory
closely follows the period $1$ orbit, which is the most
transversally unstable of the periodic orbits. Similar
observations have been previously reported for {\it two} coupled chaotic
systems \cite{pecora4}.

\begin{figure}[h]
\begin{center}
\epsfig{file = 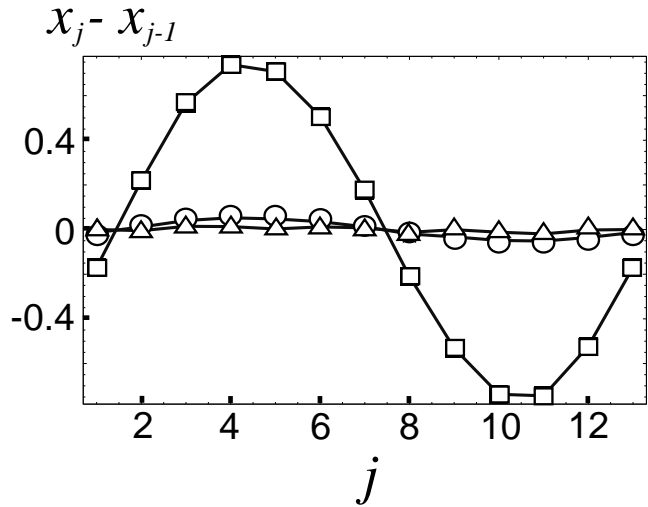, clip =  ,width=1.0\linewidth }
\caption{ $x_{j}-x_{j-1}$ versus the node index $j$ for 
 $t = 1360$ (open triangles), $t = 1385$ (open circles),
 and $t = 1410$ (open squares). Note that the burst is
absent first and grows with a long wavelength pattern.} 
\label{fig:lw}
\end{center}
\end{figure}

Finally, in Fig.~\ref{fig:lw} we plot $x_{j}-x_{j-1}$ as a
function of $j$, the oscillator index, for 
$t=1360$ (open triangles), $t = 1385$ (open circles), 
and $t = 1410$ (open squares). The desynchronization burst can be
observed developing mainly at the longest possible wavelength.

When subsequent bursts were studied in the same way, it was found
that the phase of the long wavelength burst assumed only one value.
This is due to the fact that the mismatch is `frozen', that is, 
each oscillator has a given set of parameters which differs by a given 
amount from the mean values. This fixed spatial heterogeneity 
favors certain spatial patterns over others. 
We will discuss this in more detail in Section IV.

\subsection{Short wavelength burst}

\begin{figure}[h]
\begin{center}
\epsfig{file = 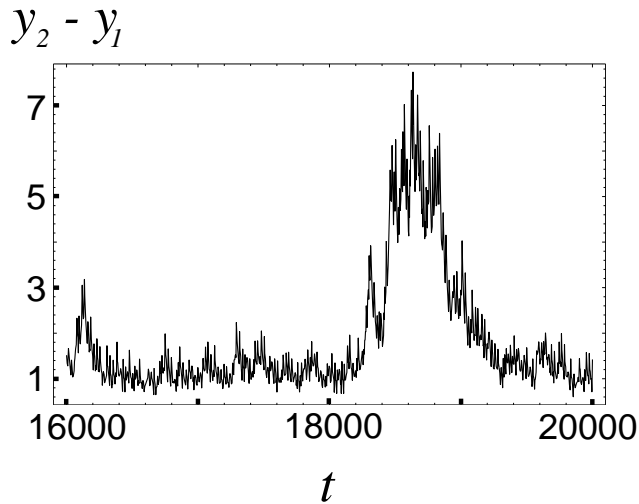, clip =  ,width=1.0\linewidth }
\caption{$y_{2} - y_{1}$ as a function of time for $8$ R\"ossler
systems  in a ring. The coupling strength $g$ was $1.09$. The
desynchronization burst develops at $t\approx 18000$, although it is
not as sharp due to the smaller magnitude of the transversal
Lyapunov exponents ($\Psi(4.36)$ in Fig.~\ref{fig:msf}).}
\label{fig:short}
\end{center}
\end{figure}

Short wavelength bursting can be expected, for example, when $N =
8$ and $g = 1.09$. In this case the value of $\lambda_{k}$
corresponding to the shortest wavelength mode yields $g\lambda_{k}
= 4.36$, which is in the bubbling region, while all the other modes are
in the stable region. In this case the observation of the bursts
is more difficult, as the transversal instability of the orbits
and the transversal stability of the attractor are less
pronounced (compare $\Psi(4.36)$ for this case vs. $\Psi(0.19)$ for the previous example 
in Fig.~\ref{fig:msf}). Accordingly, the perturbations of the parameter $a$ were
made larger, with perturbations randomly chosen with uniform
density within a $\pm6\%$ range of the ideal values of the
parameter ($a = 0.2$). In principle this is not necessary, as a burst will
eventually occur after long enough time. In practice, however, it
is necessary to reduce the waiting time to a reasonable value.
As before, the
coupled equations were solved with an initial condition on the
synchronization manifold. In Fig.~\ref{fig:short} we show $y_{2} -
y_{1}$ as a function of time for one choice of initial conditions.
The difference $y_{2} - y_{1}$ is usually positive and of
magnitude close to $1$. This asymmetry is not a surprise
since the oscillators are slightly different. For the relatively large
value of the mismatch used, this is the ``synchronized state''.
It is seen in
Fig.~\ref{fig:short} that the difference $y_{2} - y_{1}$ increases
rapidly at around $t \approx 18400$, and soon reaches values higher than $7$.
It remains large for a longer time than in the case of the long wavelength
burst (see Fig.~\ref{fig:burst}) and decays more slowly as well. 
This is in qualitative
agreement with the smaller absolute values of the master stability
functions for the short wavelength mode, both for typical orbits
on the attractor and for the periodic orbits. 
\begin{figure}[h]
\begin{center}
\epsfig{file = 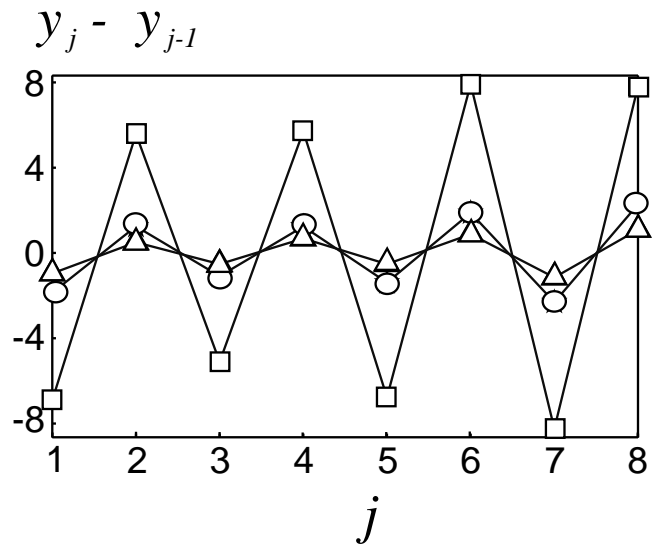, clip =  ,width=1.0\linewidth }
\caption{ $y_{j}-y_{j-1}$ versus the node index $j$ for  $t = 17970$ 
(open triangles), $t = 18270$ (open circles), and $t = 18570$
(open squares). The desynchronization burst has a short wavelength spatial pattern.}
\label{fig:short2}
\end{center}
\end{figure}

\begin{figure}[h]
\begin{center}
\epsfig{file = 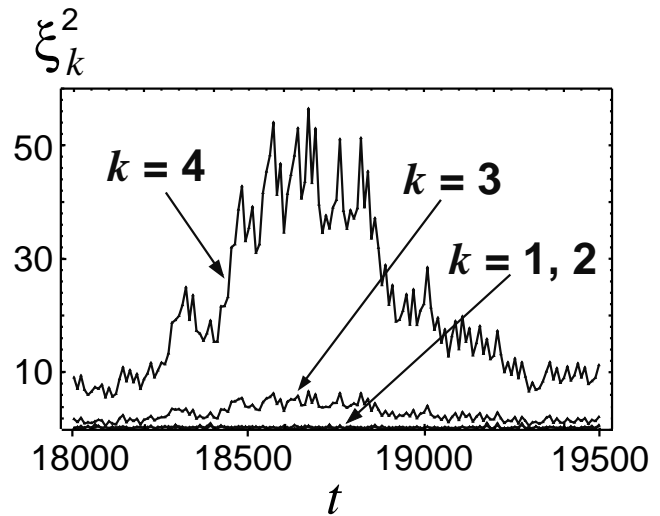, clip =  ,width=1\linewidth }
\caption{ $\xi^{2}_{k}$ as a function of time for $k = 1,2,3,4$. The shortest
wavelength component corresponds to $k = 4$ (top curve).
The second shortest one is $k = 3$ (middle curve). 
The curves corresponding to $k = 1,2$ are close to zero. }
\label{fig:short3}
\end{center}
\end{figure}

In Fig.~\ref{fig:short2} we plot $y_{j}-y_{j-1}$ as a function of
$j$, the oscillator index, for $t=17970$, $t = 18270$ and
$t = 18570$. As expected, the burst mainly affects the shortest
wavelength mode.
In order to perform a more quantitative assessment of these observations, 
we define 
$\xi_{k} = \{([\eta_{k}]_{x})^{2} +([\eta_{N-k}]_{x})^{2}\}^{\frac{1}{2}}$ 
for $1\leq k < \frac{N}{2}$ and 
$\xi_{\frac{N}{2}} = \left|[\eta_{\frac{N}{2}}]_{x}\right|$, where
$[\eta_{k}]_{x}$ is the $x$ component of the three dimensional vector
$\eta_{k}$.
The quantity $\eta_{k}$ is proportional to the magnitude  of
the vector $[\eta_{1,k}\sin\left(\frac{2\pi j k}{N}\right)+
\eta_{1,N-k}\cos\left(\frac{2\pi j k}{N}\right)]$, the 
component in the Fourier decomposition of the perturbation $\epsilon$ 
associated with modes $k$ and $N-k$. Thus,
the quantity $\xi_{k}$ represents the weigth of the modes associated to the eigenvalue 
$\lambda_{k}$, which have, for $1\leq k \leq \frac{N}{2}$, 
a wavelength of $\frac{1}{k}$'th of a full wavelength.
In Fig.~\ref{fig:short3},
we plot as a function of time the quantities $\xi^{2}_{k}$ for $k =1,2,3,4$.
The short wavelength mode ($k = 4$, upper curve) 
is dominant during the burst.

\subsection{Localized burst}

In the above examples all links had equal weights. As an example
of a case with unequal link weights we consider the case where the
previous network is modified by doubling the strength of one of
the links. Let the link whose strength is doubled be the link that
connects nodes $p$ and $p+1$. For example, for $p=4$, $N=8$, this
yields the Laplacian matrix
\begin{equation}\label{eq:laplacian}
G = \left( \begin{array}{cccccccc}
     2 & -1 & 0 & 0 & 0 & 0 & 0 & -1 \\
     -1 & 2 & -1 & 0 & 0 & 0 & 0 & 0  \\
     0 & -1 & 2 & -1  & 0 & 0 & 0 & 0  \\
     0 & 0 & -1 & 3 & -2 & 0 & 0 & 0 \\
     0 & 0 & 0 & -2 & 3 & -1 & 0 & 0  \\
     0 & 0 & 0 & 0 & -1 & 2 & -1 & 0   \\
     0 & 0 & 0 & 0 & 0 & -1 & 2 & -1    \\
     -1 & 0 & 0 & 0& 0 & 0 & -1 & 2
\end{array} \right),
\end{equation}

Adopting the analysis technique of Ref.~\cite{sott}, we can show that such
an enhanced connection has the consequence that the largest
eigenvalue of $G$ corresponds to an eigenfunction that is
exponentially localized to the region near the strong connection.
That is, for large $N$, the components of this eigenfunction decay
exponentially as the distance between the localized region and the
node corresponding to a component increases. Using the ideas of
Ref.~\cite{sott}, we now provide this analysis. The equations for the
eigenvector $w$ and eigenvalue $\lambda$ are
\begin{eqnarray}\label{eq:wp}
- 2 w_{p+1}- w_{p-1} + 3 w_{p} = \lambda w_{p},\\\nonumber
-w_{p+2}- 2 w_{p}  + 3 w_{p+1} = \lambda w_{p+1},\\\nonumber
 -w_{j-1}- w_{j+1}  + 2 w_{j} = \lambda w_{j},
\end{eqnarray}
for, respectively, nodes $p$, $p+1$ and $j$ different from $p$ or
$p+1$. Since nodes $p$ and $p+1$ are identical, we can assume, for
$k>0$, $w_{p+1 + k} = \pm w_{p-k}$. Furthermore, we propose 
the ansatz $w_{p+1+k}\propto t^{k}$, for $k>0$. 
This will be a good approximation if the mode is localized 
(i.e., $|t|<1$), and the network is big enough that $|t|^{\frac{N}{2}} \ll 1$.
In the antisymmetric case, $w_{p+1 + k} = -w_{p-k}$, 
Eqs. (\ref{eq:wp}) yield,
\begin{eqnarray}
5 - t = \lambda ,\\\nonumber
 -t - t^{-1} + 2 = \lambda
\end{eqnarray}
which gives 
\begin{equation}\label{eq:lamt}
t =-\frac{1}{3} \mbox{  , \,  } \lambda = \frac{16}{3}.
\end{equation}
Compare this eigenvalue with the largest
eigenvalue for the network in which all links have equal strength,
which has a value of $4$. The symmetric solution, $w_{p+1 + k} =
w_{p-k}$, yields $t = 1$ and $\lambda = 0$, corresponding to the
eigenvector $[1,1,\dots 1]$ of perturbations in the
synchronization manifold. 
The smallest nonzero eigenvalue remains unchanged.

\begin{figure}[h]
\begin{center}
\epsfig{file = 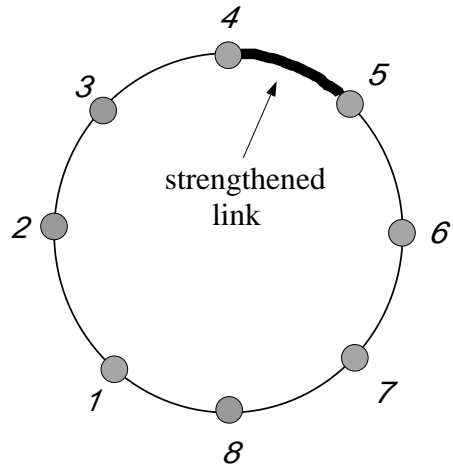, clip =  ,width=0.7\linewidth }
\caption{Arrangement of the dynamical units in a ring with the strength of the connection
between nodes $4$ and $5$ doubled. The matrix $G$ corresponding to this network is 
in Eq.~(\ref{eq:laplacian}).}
\label{fig:bolas}
\end{center}
\end{figure}

As an example, we show the localized desynchronization bursts
produced by one of these strengthened connections for the case
$N=8$, corresponding to $G$ given by (\ref{eq:laplacian}) and
the illustration in Fig.~\ref{fig:bolas}.
The parameters of the idealized system are again $a = b = 0.2$, 
and $c = 7$, with a coupling strength of $g = 0.79$.
It is remarkable that despite the small number of nodes, the
actual localized eigenvector and eigenvalue agree well with (\ref{eq:lamt})
 ($\lambda = 5.334\dots$ and $\frac{w_{6}}{w_{5}}
= -0.334\dots$).

\begin{figure}[h]
\begin{center}
\epsfig{file = 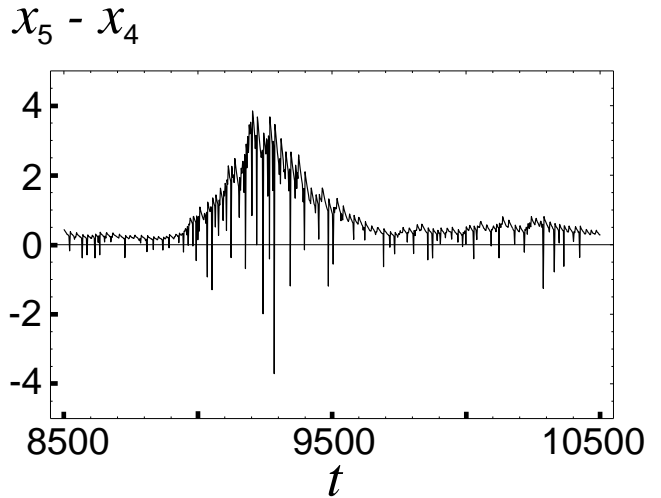, clip =  ,width=1.0\linewidth }
\caption{$x_{5}-x_{4}$ as a function of time for $N=8$ R\"{o}ssler oscillators in 
a ring with the strength of the connection between nodes $4$ and $5$ doubled. The 
coupling strength is $g = 0.79$.
A desynchronization burst starts approximately at $t \approx 9000$.}
\label{fig:local}
\end{center}
\end{figure}

In Fig.~\ref{fig:local} we show $x_{5}-x_{4}$ as a function of time.
As in the short wavelength case, the burst is not very sharp
due to the small magnitude of the transversal Lyapunov exponents.
Nevertheless, it can be seen that
the difference $x_{5}-x_{4}$ increases approximately at $t = 9000$
and returns to a relatively
small value after reaching values considerably above the average.

\begin{figure}[h]
\begin{center}
\epsfig{file = 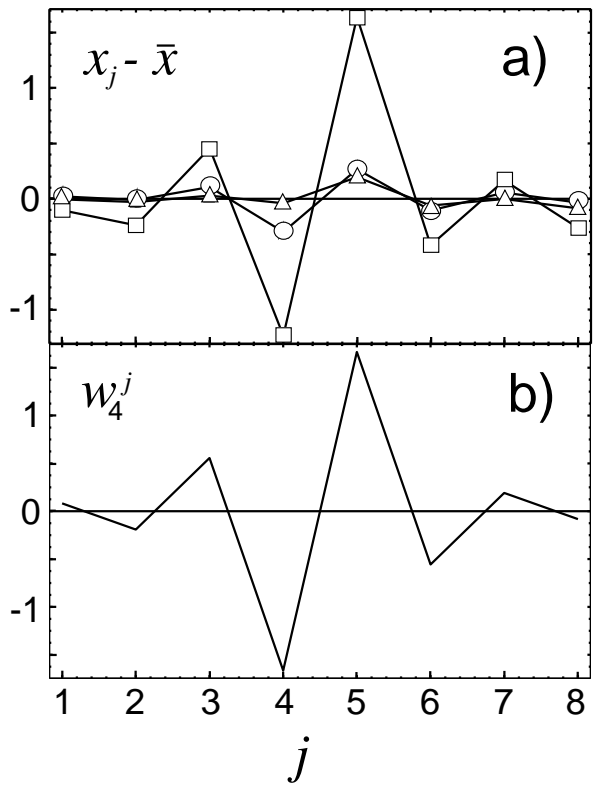, clip =  ,width=1.0\linewidth }
\caption{ a) $x_{j}- \overline{x}$ for $t = 8750$ (open triangles), 
         $t = 9000$ (open circles), 
          and $t = 9250$ (open squares), for
         the configuration in Fig.~\ref{fig:bolas}. The burst develops with the
         spatial pattern of the localized eigenvector in Fig.~\ref{fig:locajuntas}b. 
         b) Localized eigenvector of matrix $G$ in Eq.~(\ref{eq:laplacian}).}
\label{fig:locajuntas}
\end{center}
\end{figure}

In Fig. \ref{fig:locajuntas}a we plot the difference between
the $x$ coordinate of node $j$ and its mean over all nodes, $x_{j}- \overline{x}$, where 
$\overline{x} = \frac{1}{N}\sum_{j=1}^{N} x_{j}$, as a
function of the oscillator index $j$, for 
$t= 8750$ (open triangles), $t=9000$ (open circles), and $t = 9250$ (open squares). 
In Fig.~\ref{fig:locajuntas}b we show the localized
eigenvector of the Laplacian $G$ found numerically. 
As discussed before, the
desynchronization burst affects mainly nodes $4$ and $5$ (those
which share the strengthened connection) and the ones adjacent to them.
Nodes $1$,$2$,$7$ and $8$, however, maintain approximate synchronization
during the burst.

\begin{figure}[h]
\begin{center}
\epsfig{file = 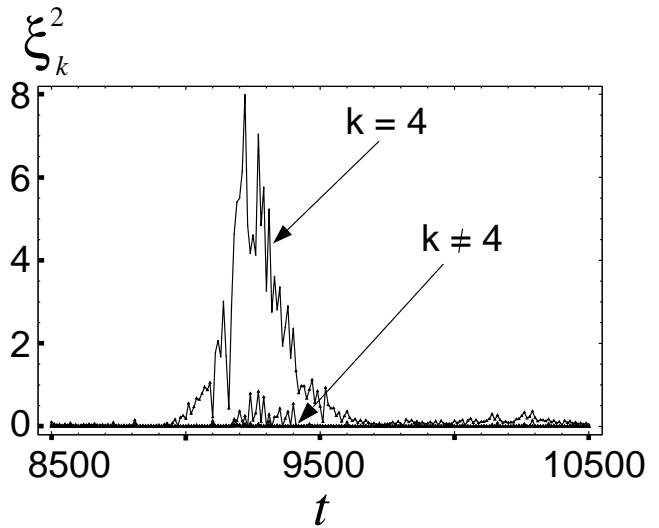, clip =  ,width=1.0\linewidth }
\caption{ $[\eta_{k}]_{x}^{2}$ as a function of time for $k =4$ 
(top curve) corresponding to the localized mode, and for $k\neq 4$ (bottom curves, close to zero),
corresponding to other modes. In the burst, the localized mode is excited first and only after
some time are the other modes also somewhat excited. The localized mode is dominant during the burst.}
\label{fig:modeweights}
\end{center}
\end{figure}

In Fig.~\ref{fig:modeweights} we show the mode weights
corresponding  to the $x$ coordinate as a function of time. The top curve
corresponds to $[\eta_{4}]_{x}^{2}$ (for the localized mode), and the curves close
to the horizontal axis to $[\eta_{k}]_{x}^{2}$, $k\neq4$, for the other modes. 
(The degeneracy of the eigenvalues is broken by the strengthened connection, 
so we do not combine $[\eta_{k}]_{x}$ and $[\eta_{N-k}]_{x}$ as before.) 
Confirming the
qualitative similarity between the eigenvector and the spatial
pattern of the desynchronization burst observed in
Fig.~\ref{fig:locajuntas}, the weight corresponding to the localized
eigenvector is seen to be dominant during the period of time in
which the burst occurs.

\section{Effects of the mismatch spatial patterns}

In this section we will discuss the effects that the mismatch 
spatial patterns have on the development of the desynchronization bursts. 
For these purposes, it will be convenient 
to rewrite Eq. (\ref{eq:coupled}) in the form

\begin{equation}\label{eq:couplednoise}
\dot{X}_{i} = \overline{F}(X_{i}) - g \sum_{j =
1}^{N}G_{ij}H(X_{j}) + Q_{i}(X_{i}),
\end{equation}
where $\overline{F}(X_{i})= F(X_{i},\overline{\mu})$ with
$\overline{\mu} = \frac{1}{N}\sum_{j=1}^{N}\mu_{j}$, and
$Q_{i}(X_{i}) = F(X_{i},\mu_{i}) - \overline{F}(X_{i})$. The term
$Q_{i}$ represents the effect of the mismatch 
and is assumed to be small. As before, we linearize around the
synchronous state to get
\begin{equation}\label{eq:linearizednoise}
\dot{\epsilon}_{i} = D\overline{F}(s)\epsilon_{i} - g \sum_{j =
1}^{N} G_{ij}DH(s)\epsilon_{j} + Q_{i}(s),
\end{equation}
where we have discarded terms of order $Q \epsilon$. With the
previous  notation and $Q=[Q_{1},Q_{2},\dots Q_{N}]$, we obtain
after the diagonalization
\begin{equation}\label{eq:componoise}
\dot{\eta}_{k} = \left(D\overline{F}(s) -  g \lambda_{k}
DH(s)\right)\eta_{k} + (QL)_{k},
\end{equation}
where $(QL)_{k}$ is the $k$'th column of the $d \times N$ matrix $QL$.
In the ring with equal coupling along each link, 
the diagonalization procedure corresponds to a
Fourier  transform. In this case, we see that the mismatch affects
the different modes according to the weigth, $(QL)_{k}$, of this particular
mode in its Fourier expansion. In other cases, for example in the
localized eigenvector, the strength of the mismatch affecting the
localized mode is proportional to the weigth of the localized
eigenvector in the eigenvector decomposition of the mismatch.
We will now discuss two applications of these results.

\subsection{Amplification of mismatch patterns when modes with the same eigenvalue burst}

We have shown that the modes of the mismatch force the corresponding modes
of the deviations from the synchronous state. When bubbling induced bursting is expected,
the size of the mismatch determines the average time between bursts \cite{bhott2}. 
Thus, the size of the mismatch component in mode $k$ determines the
average interburst time when that mode is in the bubbling regime.

When the spectrum of the matrix $G$ is degenerate, the spatial modes
of the mismatch play an extra role. All the modes sharing the 
same eigenvalue $\lambda$ have the same stability properties, and thus,
when the corresponding value $g\lambda$ is in the bubbling zone,
all eigenvectors with this eigenvalue are equally likely to appear. 
The only difference 
between these modes is the strength with which they are forced,
which is determined by the mismatch component in that mode as
shown in Eq.~(\ref{eq:componoise})
(or, if noise is present, by the noise component in that mode).

An example of this situation is the ring with connections of equal strength in the
long wavelength bursting scenario.
Since the ring is invariant with respect 
to rotations, the phase of the long wavelength oscillations can not be determined only from 
the network and dynamics part of the problem. The two 
modes with the longest wavelength (corresponding to sinusoidal and 
cosinusoidal oscillations) have the same eigenvalue.
It is the mismatch that in this case determines the phase of 
the long wavelength burst.

We will show how one can determine the phase of the
long wavelength desynchronization burst in the case of coupled
R\"ossler systems in a ring with equal coupling along each link. 
For this system, the mismatch vector $Q_{j}(X_{j})$ is given by
\begin{equation}\label{eq:q}
Q_{j}([x_{j},y_{j},z_{j}]^{T}) = \left( \begin{array}{c}
     0 \\
     y_{j} \delta a_{j}   \\
     \delta b_{j} - z_{j}\delta c_{j} \\
\end{array} \right),
\end{equation}
where $\delta a_{j} = a_{j} - \overline{a}$ und similarly 
for $\delta b_{j}$ and $\delta c_{j}$.
We define $\mathcal{F}_{k}(u)= \sum_{j=1}^{N}
u_{j} \hat{w}_{j}^{k}$, where $\hat{w}_{j}^{k}$ is the normalized $j$'th 
component of the $k$ eigenvector 
described at the beginning of Section III. With this convention,
the term $(QL)_{k}$ in equation ($\ref{eq:componoise}$) is given by
\begin{equation}\label{eq:Fq}
(QL)_{k} = \left( \begin{array}{c}
     0 \\
      y \mathcal{F}_{k}(\delta a)   \\
     \mathcal{F}_{k}(\delta b) - z \mathcal{F}_{k}(\delta c) \\
\end{array} \right).
\end{equation}
Here $\delta a = [\delta a_{1}, \delta_{2},\dots,\delta_{N}]$ and similarly for
$\delta b$, $\delta c$, and $y, z$ are the trajectories around which the linearization
was made.

We consider the case in which mismatch in one parameter
is dominant, for example $a$. The mismatch in the parameters $b$
and $c$ will be assumed negligible compared with that in $a$, so
that $\delta b$, $\delta c \ll \delta a$. In this case, only the
second component of (\ref{eq:Fq}) is of relevance. Thus
 modes $\eta_{1}$ and $\eta_{N-1}$ are excited
with a strength proportional, respectively, to
$\mathcal{F}_{1}(\delta a)$ and $\mathcal{F}_{N-1}(\delta a)$; see
(\ref{eq:componoise}). The magnitude of $\eta_{k}$
will be proportional to $\mathcal{F}_{k}(\delta a)$, and thus the 
excitation of the long
wavelength mode (which is the only one for which perturbations grow) 
is proportional to 
\begin{equation}
\mathcal{F}_{1}(\delta a)\sin\left(\frac{2\pi j}{N}\right)+
\mathcal{F}_{N-1}(\delta a) \cos\left(\frac{2\pi j}{N}\right)
\end{equation}
\begin{equation}
\propto \sin\left(\frac{2\pi j}{N} + \phi \right) ,\\
\end{equation}
where $\tan \phi = \mathcal{F}_{N-1}(\delta a)/\mathcal{F}_{1}(\delta a)$.

\begin{figure}[h]
\begin{center}
\epsfig{file = 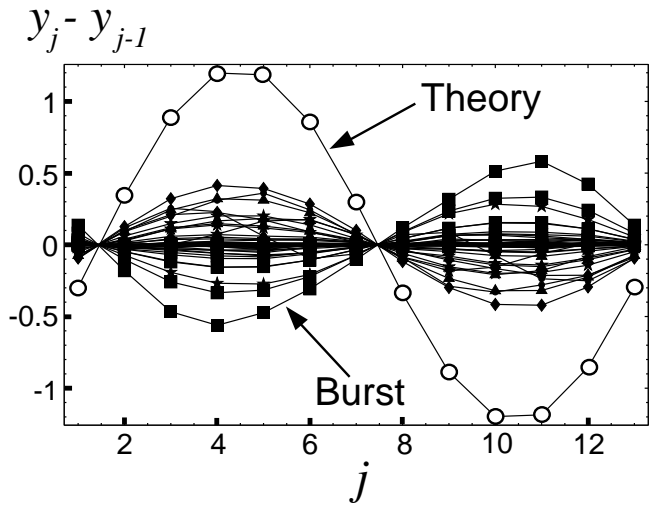, clip =  ,width = 1.0\linewidth }
\caption{$y_{j}-y_{j-1}$ for different times during a burst (filled symbols), and
a scaled version of $\sin\left(\frac{2\pi
j}{12}+\phi\right)-\sin\left(\frac{2\pi (j-1)}{12}+\phi\right)$ with $\phi$ 
as given in the text
(open circles). The phase of the burst spatial pattern coincides with the 
phase of the 
long wavelength component of the mismatch.
} \label{fig:noisepattern}
\end{center}
\end{figure}

We now show results of numerical simulations illustrating the above. 
The parameters $N$ and $g$ will be as in the
long wavelength example in the previous section. We use the same
random set of perturbations used in that example. 
As described above, we obtained the phase
$\phi$ of the long wavelength component of the vector $\delta
a_{i}$. In Fig.~\ref{fig:noisepattern} we plot $y_{j}-y_{j-1}$ for
different times during a burst (filled symbols). In the
same Figure, we plot a scaled version of $\sin\left(\frac{2\pi
j}{12}+\phi\right)-\sin\left(\frac{2\pi (j-1)}{12}+\phi\right)$
(open circles). The phase of the desynchronization burst
is in agreement with that of the long wavelength component of the
mismatch.

When the mismatch affects predominantly one parameter as in this case, the
phase of the bursts can be predicted as described
above. When mismatch in different parameters is comparable, the phases of the
long wavelength modes of the different parameter mismatches
compete and the bursts develop with one of these phases or with a 
combination of them. 

It must be emphasized that this analysis is possible only when there
is a degeneracy of the eigenvalues. For example, the location of the localized burst
can not be determined in this way, as it is fixed in the position of the strengthened link. 
In this case, the mismatch component in the localized mode 
would only affect the average time between bursts.

\subsection{Artificial supression of unstable modes using knowledge of the mismatch}

We will now discuss another consequence of Eq.~(\ref{eq:componoise}). 
We imagine a situation where we are given a number of nearly identical 
oscillators that we are to connect in a network which we desire to be in synchronism 
as much as possible. Furthermore, we imagine that, through measurements made
individually on each oscillator, we are aware of the amount of mismatch
in each oscillator. The question we address is this: Using our knowledge of the
individual mismatches, how should we arrange the oscillators in the network
so as to best supress the frequency of desynchronism bursts? To answer this question,
we note that, according to the previous discussion, we should reduce the mismatch 
component in the mode which is in the bubbling region.
Since the size of the mismatch affects the average interburst time
\cite{bhott2}, reducing this component is desirable if one wants to 
improve the quality of the synchronization. This can be accomplished by judiciously arranging 
the dynamical units so that the $k$'th mode of the mismatch is minimized when
the corresponding value $g \lambda_{k}$ is 
in the bubbling region. For example, to supress long wavelength bursts, one may 
arrange the units so that the parameter errors alternate above and below the mean. 
To supress the localized bursting described in the previous section, one 
could arrange the
units so that those with the more similar parameters are the ones in the region of
the strengthened connection.

\begin{figure}[h]
\begin{center}
\epsfig{file = 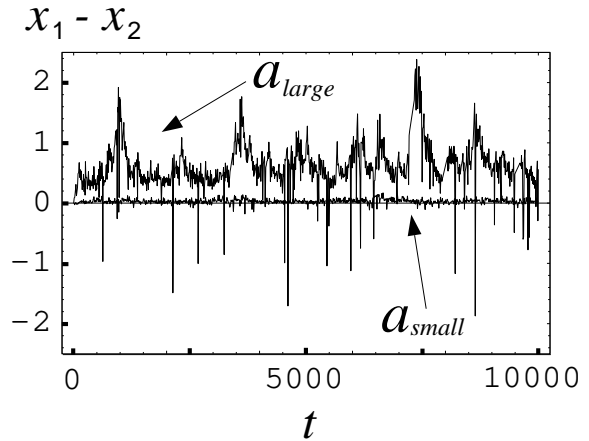, clip =  ,width=0.9\linewidth }
\caption{ $x_{1}-x_{2}$ as a function of time for a configuration
of oscillators with a large (top  curve) and with 
a small (curve closer to zero) short wavelength component of the mismatch. 
The quality of the synchronization is much better in the second case.}
\label{fig:difere}
\end{center}
\end{figure}

As a concrete example, we test this idea using simulations for the case of short 
wavelength bursting presented in the previous Section. We again assume for simplicity
that mismatch in the parameter $a$ is dominant. We generate random perturbations in
the parameter $a$ within a $\pm 6 \%$ range of the value $a = 0.2$, as explained in the 
previous section. 
With this set of parameters given, we set up the dynamical units in the ring 
using two different permutations of their positions. 
One of them ($a_{s}$) has a smaller and the other ($a_{l}$) a larger short
wavelength component $\mathcal{F}_{4}(a)$ than the original random sequence. The ratio 
$\mathcal{F}_{4}(a_{l})/\mathcal{F}_{4}(a_{s})$ is approximately $15$.
In Fig~\ref{fig:difere} we plot $x_{1}-x_{2}$ as a function of time for configuration
$a_{large}$ (top curve) and for configuration $a_{small}$ (bottom curve). 
The difference $x_{1}-x_{2}$ is much smaller in the former case than in the 
latter, roughly
by a factor of $15$, as can be expected from the ratio 
$\mathcal{F}_{4}(a_{large})/\mathcal{F}_{4}(a_{small})$. 
This qualitative example illustrates how one can use knowledge of the 
mismatch to supress undesired instabilities.

\section{Spatial patterns of deviations from the stable synchronous state}

So far, we have concentrated in the case in which the value of $g\lambda_{k}$ is in
the bubbling regime for one mode $k$ and in the stable regime for the other modes, so 
that desynchronization bursts occur sporadically. As we have seen, these bursts present
spatial patterns on the network.

If synchronization is desired, one would might try to avoid the bubbling
regime by designing the network and adjusting the coupling strength so that all the
modes lie in the stable zone. One would also strive to reduce the mismatch, but as 
mentioned before, there are practical limitations on how much one can 
make the oscillators exactly the same.

If $\Psi(g\lambda_{k})$ is negative for all
modes (indicating transversal stability of the synchronous state) 
one can have, depending on the degree of transversal stability, 
fair synchronization even with relatively
large amounts of mismatch. If one is to operate under such conditions,
it is important to know the characteristics of the deviations from the synchronous state.

Thus we ask in this scenario: 
How large are the spatial patterns of the
deviations from the synchronous state, and how does this depend on the mismatch
and on the degree of transversal stability?

The spatial modes of these deviations obey Eq. (\ref{eq:componoise}). In the absence
of the term $(QL)_{k}$, the zero solution is stable, and typical perturbations from it
decay, having a negative Lyapunov exponent given by $h_{k} \equiv \Psi(g \lambda_{k})$.
The first term in the right hand side of Eq.~(\ref{eq:componoise}) 
can be thougt of as a damping term with 
a damping rate given by $h_{k}$, and the second term, $(QL)_{k}$, as a forcing term. Since we are
considering the stable case, these two factors, on average, cancel each other. By definition,
the Lyapunov exponent for the system without mismatch is given by 
$h_{k}=\langle\frac{\eta_k^T(D\overline{F}-g\lambda_k DH)\eta_k}{\left|\eta_k\right|^2}\rangle$, where the angle
brackets indicate time average. Assuming a solution $\eta_k$ of the system with mismatch to yield the same 
value of this time average, we left multiply
 Eq.~(\ref{eq:componoise}) by $\eta_{k}^{T}\left|\eta_{k}\right|^{-2}$
and average to obtain
\begin{equation}\label{eq:exact}
\left|h_{k}\right| \approx 
 \langle\frac{\eta_{k}^{T}(QL)_{k}}{\left|\eta_{k}\right|^{2}}\rangle\sim
\langle\frac{\left|(QL)_{k}\right|}{\left|\eta_{k}\right|}\rangle,
\end{equation}
where the angle brackets indicate time average.
This leads to the following rough estimate,
\begin{equation}\label{eq:machete}
\langle\left|\eta_{k}\right|\rangle 
\sim \frac{\langle \left|(QL)_{k}\right|\rangle}{\left|h_{k}\right|}
\end{equation}

As an example we consider R\"{o}ssler units in a ring with all
connections of equal strength. We choose $N = 8$, $g = 0.6$ 
($\Psi(g\lambda_{k})<0$ for all values of $k$).
Furthermore, we add a random perturbation to the parameter $a$ of each
oscillator chosen uniformly from within a $\pm0.1\%$ range of $a = 0.2$.

\begin{figure}[h]
\begin{center}
\epsfig{file = 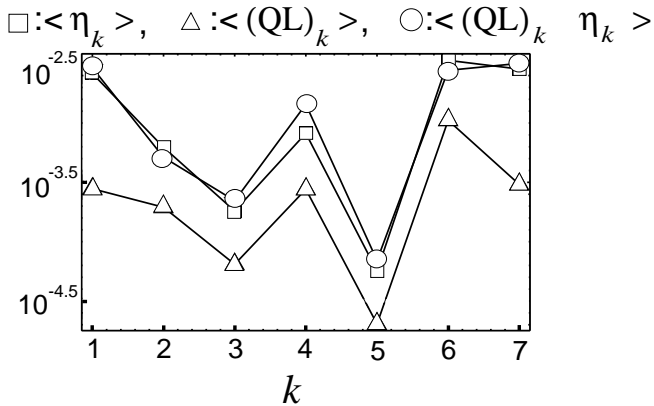, clip =  ,width=1.0\linewidth }
\caption{$\langle\left|\eta_{k}\right|\rangle$ (open squares),  
$\langle\left|(Q L)_{k}\right|\rangle$ (open triangles), 
and  $\frac{\langle\left|(Q L)_{k}\right|\rangle}{h_{k}}$ 
(open circles) for $N = 8$, $g = 0.6$, $k = 1,\dots,7$. 
The forcing term (open triangles) roughly determines the response (open squares).
The corrected forcing term (open circles) matches well the response (open squares). }
\label{fig:misco}
\end{center}
\end{figure}

In Fig.~\ref{fig:misco} we show, for $k = 1,\dots,7$, the quantities 
$\langle\left|\eta_{k}\right|\rangle$ (squares), 
$\langle\left|(Q L)_{k}\right|\rangle$ (triangles), 
and $\frac{\langle\left|(Q L)_{k}\right|\rangle}{\left|h_{k}\right|}$ (stars).
The magnitudes of the forcing term 
for the different modes ($\langle\left|(Q L)_{k}\right|\rangle$) span roughly two orders of magnitude, 
and the magnitude of the response ($\langle\left|\eta_{k}\right|\rangle$) looks roughly 
proportional to the latter. When the forcing term is corrected by dividing it by the magnitude
of the corresponding
Lyapunov vector $\left|h_{k}\right|$, the resulting quantity 
($\frac{\langle\left|(Q L)_{k}\right|\rangle}{\left|h_{k}\right|}$) 
matches very well the observed response.

\section{Conclusions}

We have studied the stability properties of the synchronized state in a network of
coupled chaotic dynamical units when these have a small heterogeneity.
We have shown that when the dynamical units that are coupled in a network are sligthly
different, the synchronized state can be interrupted by large infrequent 
desynchronization bursts for some values of the parameters. 
The range of the parameters for which this phenomenon is expected can be obtained by
performing a master stability function analysis of the chaotic attractor and of the 
periodic orbits embedded in it.

The desynchronization bursts are induced by the bubbling phenomenon, 
and have spatial patterns on the network. These spatial patterns can be predicted
from the eigenvectors of the Laplacian matrix $G$ and the master stability functions 
mentioned above. We showed examples illustrating the development of bursts with 
spatial patterns. One of our examples showed that the strengthening of a single connection
might destabilize the nodes near this connection, while leaving the rest
of the network approximately synchronized.

Direct measurement of the parameter mismatch in the elements of a network might prove useful.
We discussed how this knowledge could be used to reduce the frequency of bursts and to predict the 
relative weights of different spatial patterns in a burst.
We also discussed how one could, from knowledge of the mismatch and of the master 
stability function, describe the spatial patterns and magnitude of the deviations 
from the synchronized state when the synchronization of the corresponding identical unit
system is robust.

We emphasize that although we did not discuss the effects of noise, the phenomenon described
in this paper also occurs for noisy identical oscillators. Desynchronization bursts with
spatial patterns are expected for noisy, identical oscillators if one has them for
noiseless, nonidentical oscillators. The difference is that the parameter mismatch is always
`frozen', in the sense that the mismatch is always the same for each oscillator, whereas 
for noise this is not the case.

Acknowledgements: This work was sponsored by ONR (Physics) and by
NSF (contracts PHYS 0098632 and DMS 0104087).

\end{document}